\documentstyle[multicol,aps,epsfig,bbox]{revtex}
\begin{document} 
\newcommand{\beq}{\begin{equation}} 
\newcommand{\eeq}[1]{\label{#1} \end{equation}}  
\newcommand{\half}{{\textstyle\frac{1}{2}}}  
\newcommand{\ee}{\end{equation}} 
\newcommand{\bea}{\begin{eqnarray}}  
\newcommand{\eea}{\end{eqnarray}} 
\newcommand{\beqar}{\begin{eqnarray}}  
\newcommand{\eeqar}[1]{\label{#1} \end{eqnarray}}  
\newcommand{\gton}{\mathrel{\lower.5ex \hbox{$\stackrel{> }
 {\scriptstyle \sim}$}}}
\newcommand{\lton}{\mathrel{\lower.5ex \hbox{$\stackrel{< }
 {\scriptstyle \sim}$}}}

\title{Jet Quenching and  the 
       $\bar{p} \gton \pi^-$  Anomaly at RHIC}  
 
\author{Ivan~Vitev$^{1}$ and Miklos~Gyulassy$^{1,2}$} 
 
\address{$^1$ Department of Physics, Columbia University,  
         538 W. 120-th Street, New York, NY 10027, USA\\ 
         $^2$ Collegium Budapest, Szentharomsag u.2,
H-1014 Budapest, Hungary } 
 
\maketitle 
 
\begin{abstract} 
PHENIX data on $Au+Au$ at $\sqrt{s}= 130$~AGeV  suggest that 
$\bar{p}$ yields may exceed $\pi^-$ at high $p_{\rm T} > 2$~GeV/c.
We propose that jet quenching in central collisions suppresses the 
hard PQCD component of the spectra in central $A+A$ reactions,
thereby exposing a  novel 
component of baryon  
dynamics that  we attribute to (gluonic) baryon junctions. 
We predict that the observed   $\bar{p} \gton  \pi^-$ and the $p>\pi^+$ 
anomaly at $p_{\rm T} \sim 2$ GeV/c is 
limited  to a finite $p_{\rm T}$ window that decreases  with increasing 
impact parameter.
\vspace{.2cm} 
 
\noindent {\em PACS numbers:} 12.38.Mh; 24.85.+p; 25.75.-q  
\end{abstract} 

\begin{multicols}{2}

{\bf Introduction.} Recent data 
on  $Au+Au$ reactions at $\sqrt{s}=130$~AGeV from 
the Relativistic Heavy Ion Collider (RHIC)
have revealed a number of 
new phenomena at moderate high $p_{\rm T} \sim 2-6$~GeV/c.
The  high $p_{\rm T}$ spectra of $\pi^0$ in  central
collisions were found by PHENIX~\cite{phenix} to be suppressed 
by a factor $\sim 3-4$ relative to perturbative QCD (PQCD) 
predictions scaled by the Glauber binary collision density  
($T_{AA}({\bf b})$). Also in non-central collisions, STAR~\cite{star} 
found that the linearly increasing  azimuthal asymmetry moment, 
$v_2(p_{\rm T})$, saturates at $\sim 0.15$ 
for $p_{\rm T}>$ 2 GeV/c 
in contrast to predictions based on ideal hydrodynamics~\cite{kolb}. 
From Ref.~\cite{gvwprl} these features were interpreted as 
evidence  for jet quenching in a dense gluon plasma with 
rapidity density $dN^g/dy\sim 1000$. The initial proper gluon density  
may thus have reached far into the deconfined QCD phase with 
$\rho_g(\tau\sim 0.2\; {\rm fm/c}) > 10/{\rm fm}^3$.
Jet quenching at RHIC, if confirmed by further measurements, 
opens the door to a 
new class of diagnostic tools that probe $AA$ dynamics and 
the transient quark-gluon plasma created in such collisions.

One of the  unexpected results reported by PHENIX and STAR 
is that in contrast to the strong $\pi^0$ quenching at $p_{\rm T}>2$ GeV/c, 
the summed negatively charged ($\pi^-+K^-+\bar{p}$) and the 
corresponding positively charged hadron spectra were found to be  
quenched by only a  factor $\sim 2$~\cite{phenix,pt}. Even more  
surprisingly, the identified particle 
spectra analysis at PHENIX~\cite{julia} suggests that 
$R_B(p_{\rm T})=\bar{p}/\pi^- \gton 1$ at $p_{\rm T}>2$~GeV/c. 
Thus, baryon and anti-baryon production may in fact  dominate the 
moderate high $p_{\rm T}$  hadron flavor yields, a phenomenon never before
observed.

These and other data point to novel 
baryon 
transport dynamics  playing role in nucleus-nucleus ($AA$) reactions.
An important indicator of this are STAR~\cite{nuqm01} data
that revealed a  high valence proton rapidity density 
$(\sim 10)$,
five units from the fragmentation regions, and a $\bar{p}/p\simeq 0.65$
at mid-rapidity. An attractive dynamical model that explains
copious mid-rapidity baryon and anti-baryon production  is based on the 
existence of topological gluon field configurations 
(baryon-junctions)~\cite{junction,dima96,gvlong}.
Junctions predict long range baryon number transport in 
rapidity as well as  hyperon enhancement (including $\Omega^-$)~\cite{dima96}
and considerable $p_{\rm T}$ enhancement~\cite{junction,gvlong} relative 
to conventional diquark-quark string fragmentation~\cite{quench}.
In this paper we propose that the baryon/meson anomaly
is due to the interplay between the jet quenched~\cite{gvwprl,quench,glv}
hard component ($dN_h$)
and the phenomenological 
 soft to moderate $p_{\rm T}$ component ($dN_s$) of 
$\bar{p}$ and $\pi^-$.
The observed 
baryon junction  component is assumed to
exist in nucleon-nucleon ($NN$) and well as $AA$ reactions, 
but in $NN$~\cite{junction,dima96} it has a smaller amplitude, and its 
contribution to the high  $p_{\rm T}$ baryon spectrum  is ``obscured'' 
by unquenched mini-jet fragmentation into pions.

In this paper we extend the study of $K^\pm/\pi^\pm$~\cite{levai} 
and $p/\bar{p}$~\cite{wang}  ratios to 
compute the expected differential baryon flavor ratio,
$R_B(p_{\rm T}) \equiv \bar{p}/\pi^-$.
We also generalize the two component soft+hard dynamical model
of~\cite{gvwprl} to include the novel 
baryon junction component.
We predict that the enhancement of the 
$\bar{p}/\pi^-$ ratio reported  by PHENIX is actually limited to a finite 
moderate $p_{\rm T}$ range 2-6~GeV/c. Beyond this range, the hadron ratios are 
expected  to converge back to the quenched PQCD base.
In $Au+Au$, our proposed mechanism  can be further tested  
via a predicted systematic decrease of $\bar{p}/\pi^-$ 
with increasing impact parameter,
i.e. decreasing nucleon participant number.

{\bf Reference  PQCD flavor composition.}
The standard PQCD approach 
expresses the differential hadron cross section in $NN \rightarrow hX$ 
as a convolution of the measured structure functions 
$f_{\alpha/N}(x_\alpha,Q_\alpha^2)$  for the interacting partons 
($\alpha = a,b$), with the fragmentation function 
$D_{h/c}(z,Q^2_c)$ for
the leading scattered parton $c$ into a hadron of flavor $h$ and the
elementary parton-parton cross sections 
$d\sigma^{(ab \rightarrow cd)}/d\hat{t}$.
Folding over an intrinsic  partons  (Gaussian) $f({\bf k}_{\rm T})$
distribution: 
\begin{eqnarray}
E_{h}\frac{d\sigma^{NN}}{d^3p} &=&
K   \sum_{abcd} \int\! dz_c dx_a 
dx_b \int d^2{\bf k}_{{\rm T}a} d^2{\bf k}_{{\rm T}b} \nonumber \\[1ex] 
&\;& f({\bf k}_{{\rm T}a})f({\bf k}_{{\rm T}b})
f_{a/p}(x_a,Q^2_a) f_{b/p}(x_b,Q^2_b) \nonumber \\[1ex]
&\;&  D_{h/c}(z_c,{Q}_c^2) 
 \frac{\hat{s}}{\pi z^2_c} \frac{d\sigma^{(ab\rightarrow cd)}}
{d{\hat t}} \delta(\hat{s}+\hat{u}+\hat{t}) \; ,
\label{hcrossec}
\end{eqnarray}
where $x_a, x_b$ are the initial  momentum  fractions  carried 
by the interacting partons, $z_c=p_h/p_c$ is the momentum fraction carried 
by the observed hadron. For $NN$  we use phenomenological smearing 
$\left\langle {\bf k}_{\rm T}^2 \right \rangle = 1.8$~GeV$^2$/c$^2$.
Comparison between the PQCD calculation~(\ref{hcrossec}) and the 
charged hadron multiplicities measured by the ISR ($pp$)  and UA1 
($\bar{p}p$) experiments~\cite{isrua1} at 
$\sqrt{s}=53$~GeV and $200,900$~GeV  respectively  are in good 
agreement~\cite{gvlong} for various structure and fragmentation function 
parameterizations~\cite{struct,fragm}. 
\begin{center}
\vspace*{8.9cm}
\includegraphics{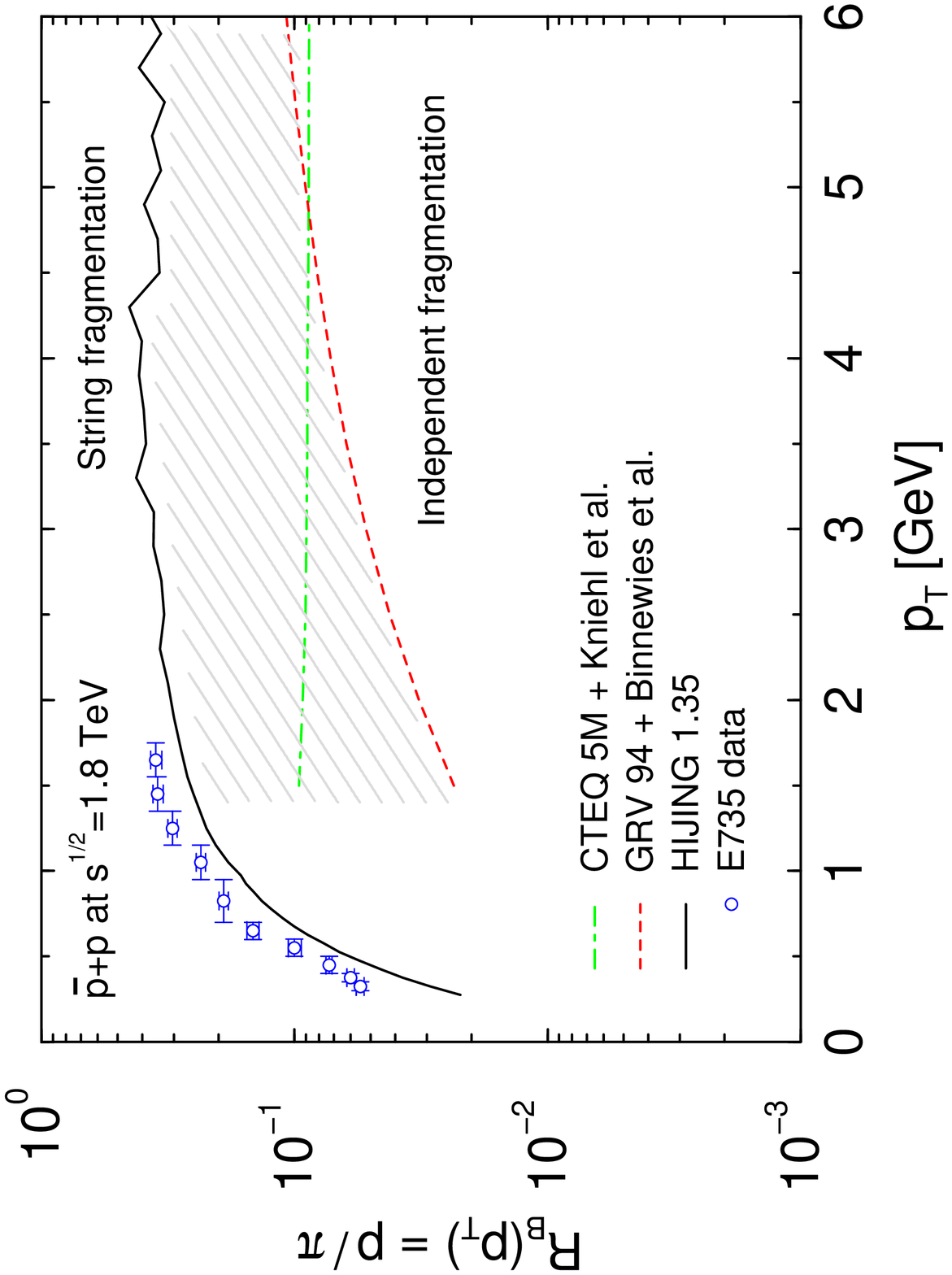}
\vspace{-3.1cm}
\end{center}
\begin{center}
\begin{minipage}[t]{8.5cm}
      { FIG. 1.} {\small  The ratio $R_B \approx
(p+\bar{p})/(\pi^++\pi^-)$  is shown
for $p\bar{p}$ at $\sqrt{s}=1.8$~TeV as a function of $p_{\rm T}$.
Low $p_{\rm T}$ FNAL E735 data~\protect{\cite{e735}} 
is shown for comparison. 
String fragmentation of mini-jets via HIJING1.35~\protect{\cite{quench}} 
is also shown.} 
\end{minipage}
\end{center}

Fig.~1 compares $R_B(p_{\rm T})$ from  the conventional independent 
and string fragmentation PQCD phenomenologies to available E735~\cite{e735}  
data from $\sqrt{s} = 1.8$~TeV $\bar{p}p$ collisions.  
While the pion dominated charged particle inclusive
data are fit  well by~(\ref{hcrossec}) with different fragmentation
models~\cite{gvlong}, there is a large difference between
the predicted $\bar{p}/\pi^-$ ratios in the shown $p_{\rm T}$ range.
Comparing results from two different parametrizations
of fragmentation functions~\cite{fragm} 
and the HIJING1.35 string fragmentation event generator~\cite{quench} 
reveals a rather large theoretical 
uncertainty  $\sim 3$ on $R_B$ at high $p_{\rm T}$.
Nevertheless, the prediction that $R_B \lton 0.5$ at high $p_{\rm T}$
is robust and intuitively reasonable given that both quark and
gluon jets prefer to fragment into lighter mesons than baryons.
We conclude that a systematic study of the $p_{\rm T}$
dependence of the baryon/meson ratios at RHIC  should include the 
$\sim 3$ uncertainty of the baryon fragmentation schemes shown 
by an error band in Fig.~1, but the PQCD predictions for 
$\bar{p}/\pi^-$ is that it stays below 0.5 for 
$p_{\rm T}> 2$~GeV/c.

{\bf Anomalous 
baryon component.}
We generalize next the (soft+hard) two component model 
in Ref.~\cite{gvwprl} and test it against the observed
charged particle quenching pattern~\cite{phenix,star}.

In $AA$ collisions the semi-hard PQCD component is 
reduced due to the non-abelian energy 
loss of hard partons before fragmentation as in~\cite{gvwprl}. 
In the calculations below, we use radiative spectrum  $dI/dx$  
(corresponding to fractional energy loss  $\Delta E/E$) computed 
in the GLV (Gyulassy-L\`evai-Vitev)  formalism~\cite{glv}. 
The numerical techniques developed in~\cite{glv} allow us to extend the 
jet quenching computation to partonic energies  as low as $\sim$5~GeV that
are relevant for the $p_{\rm T}$ range addressed in this paper.   
Jet production and propagation through the medium  is treated as 
in~\cite{gvwprl} including the realistic Wood-Saxon nuclear geometry. 
We also include 1+1D Bjorken and 1+3D Bjorken+transverse expansion of 
the interaction region.
In~\cite{gvw2} the {\em azimuthally averaged} energy loss was found 
both analytically and numerically to have little
sensitivity to transverse flow 
\begin{equation}
\left\langle \Delta E_{\rm 3D} \right \rangle_\phi  \approx   
\left\langle   \Delta E_{\rm 1D} \right \rangle_\phi  \;\;.
\label{1d3dfac}
\end{equation}  
In the computation of the PQCD component we take into account
nuclear shadowing as in~\cite{quench} and Cronin effect as in 
Ref.~\cite{cronin}. In addition, we take into account multi-gluon
fluctuations of the energy loss via the opacity 
renormalization approach as discussed in \cite{fluc}.
Fluctuations in the GLV approach effectively reduce
the opacity $\chi\propto dN^g/dy$ by a factor $\sim 0.5$.
The fitted initial gluon density quoted in Fig.~2 below takes that
reduced energy loss effect into account.

We parameterize the  soft phenomenological component of moderate 
$p_{\rm T}$ hadrons as follows:
\begin{equation}
\frac{dN_s ({\bf b})}{dyd^2{\bf p}_{\rm T}} = 
\sum\limits_{\alpha=\pi,K,p,\cdots}
\frac{dn^\alpha}{dy}({\bf b})\;\frac{e^{-p_{\rm T}/T^\alpha({\bf b})}}
{2\pi (T^{\alpha}({\bf b}))^2}
\; . \label{flavslope}
\end{equation}
As in string models the soft  component is assumed 
to scale with the number of participants ($N_{part}$). 
In Eq.~(\ref{flavslope}) we also account for the possibly different 
mean inverse slopes $T^\alpha$ for baryons and mesons. 
In the junction picture~\cite{junction}, 
the large $T^{\bar{p}}$ may arise from 
the predicted~\cite{dima96} smaller junction trajectory slope
$\alpha_J^\prime\approx \alpha_R^\prime/3$.  
This  implies that the {\em effective} string tension
is  three times higher than $1/(2\pi\alpha_R^\prime)\approx 1$ GeV/fm
leading in the massless limit to 
$\langle p_{\rm T}^2\rangle_J \simeq 
3 \, \langle p_{\rm T}^2 \rangle_R$.
In terms of the string model the factor three enhancement of the
mean square $p_{\rm T}$ is due to the random walk in 
$p_{\rm T}$ arising from the decay of the three strings attached to 
the junction. Naively, we would thus expect
$T^{\bar{p}} \simeq  \sqrt{3}\, T^{\pi^-}$. In a detailed Monte Carlo
study~\cite{Topor02} using HIJING/${\rm B\bar{B}}$~\cite{dima96}
this relation was indeed found to hold in the $1\le p_{\rm T}\le 2$~GeV/c
range with $T^\pi\approx 220$ MeV and $T^p\approx 370\; 
{\rm MeV}\;  \sim \sqrt{3}\, T^\pi$
with junctions included
and $T^p\sim 270$~MeV without. At higher $p_{\rm T}$
the PQCD mini-jets cause the apparent pion slope to
increase systematically with $p_{\rm T}$, while at lower $p_{\rm T}$ 
the resonances cause  $T^\pi$ to be smaller.
In going to peripheral collisions a small decrease
in the mean inverse slopes $\Delta T^{\pi^-} = -10$~MeV and 
$\Delta T^{\bar{p}} = -50$~MeV for approximate consistency 
with $\bar{p}p$ data~\cite{e735} is introduced.

The general hydrodynamic solution~\cite{rischke} including
transverse flow is of course  much more complex at low $p_{\rm T}$ 
than our simplified parameterization~(\ref{flavslope}). However, at
large $p_{\rm T}$ all hadronic inverse transverse slopes tend to a 
Doppler shifted value ($T_{eff}=T_f\exp(\eta_r)$, in terms of 
the transverse flow rapidity $\eta_r$ at a freeze-out isotherm $T_f$). 
As we show below, boosted thermal sources 
including relativistic transverse flow does 
not predict an anti-baryon anomaly in the 
$p_{\rm T}\sim 2$~GeV/c range~\cite{gvlong}.

The baryon transport mechanism suggested in~\cite{dima96} predicts
that the valance baryon number per unit rapidity has the form
\begin{equation}
\frac{dN^B}{dy} \simeq  \frac{dN^p}{dy} - \frac{dN^{\bar p}}{dy} =
\beta Z \frac{\cosh (1-\alpha_B(0))y}
{\sinh (1-\alpha_B(0))Y_{\rm max} } \;\;,  
\label{baryotrans}
\end{equation}
where $\alpha_B(0)\simeq 1/2$ is the baryon junction exchange intercept.
For $Au+Au$ reactions  at $\sqrt{s}=130$~AGeV Eq.~(\ref{baryotrans}) 
predicts $dN^B/dy \simeq 10$ which is in good agreement with 
RHIC data~\cite{nuqm01}.

Our phenomenological soft string and baryon junction components depends
on 2 parameters - the total charged particle rapidity density 
and the mean inverse slope for pions. We take $dN^{ch}/dy \simeq 650$ for 
the top 6\% central events from experiment~\cite{phobos} 
and $T^{\pi^-}\simeq 220$~MeV. It has been argued that the soft string 
and baryon junction dynamics is manifested in  nucleon-nucleon
collisions~\cite{junction,dima96}. Therefore the existing $pp$
data~\cite{e735,exdata} provide us  with reference integrated particle 
ratios and predictions  (e.g. $\bar{p}/\pi^- \simeq 0.06$) for the 
case of $AA$ reactions. These predictions are found to be consistent 
with the measurements at RHIC within 30\%.
Deviations of those $p_{\rm T}$ integrated ratios 
from the $\bar{p}p$ case are expected 
(e.g. strangeness enhancement) but they cannot account for the  factor
of $\sim 5$ anomalous enhancement of the $p_{\rm T}$ differential 
baryon/meson ratio reported  by PHENIX~\cite{julia}.

The full hadronic spectrum is a sum of soft and hard components
\beq
\frac{dN ({\bf b})}{dyd^2{\bf p}_{\rm T}} =
\frac{dN_s ({\bf b})}{dyd^2{\bf p}_{\rm T}}  + 
\frac{dN_h ({\bf b})}{dyd^2{\bf p}_{\rm T}}  \;\;,
\eeq{hadtot}
with a hard component computed from the cross section in nucleon-nucleon 
collisions modified by the medium induced  energy 
loss~\cite{gvwprl,glv,levai} and the Glauber profile density at a 
given impact parameter. 
To avoid over-counting, we smoothly turn on 
the power law  PQCD component at
$p_{\rm T}\sim 2$~GeV/c~\cite{gvwprl}.

We first compute the ratio $R_{AA}(p_{\rm T})$ of the hadron 
multiplicity in $Au+Au$ at $\sqrt{s}=130$~AGeV
normalized to the corresponding  multiplicity in $\bar{p}{p}$
reactions scaled by the number of binary collisions. Such scaling 
is relevant for the comparison of the moderate to high $p_{\rm T}$ parts
of the hadronic spectra and in absence of nuclear effects is expected to
be unity. Fig.~2 shows $R_{AA}$  in the 10\% central 
collisions for neutral pions and inclusive charged hadrons. 
We note that including only nuclear shadowing and Cronin effect gives a
ratio consistent with unity within 10\% for $p_{\rm T}\geq 2$~GeV/c and
inconsistent with data if jet 
quenching is not taken into account~\cite{cern}.
The fitted parton energy  loss is specified by initial gluon 
rapidity density $dN^g/dy=800\pm 100$ and this density is close
to predictions based on saturation models~\cite{ekrt}. 
This high gluon density takes into
account the reduction of the effect of induced radiative energy loss
due to multi-gluon fluctuations as discussed in detail in 
Ref.~\cite{fluc}.

In relativistic 
heavy ion collisions early thermalization of the soft background partons 
is expected. This allows the estimate of the Debye 
screening  $\mu$ (which is a natural 
infrared cut-off in medium) from thermal pQCD.
The screening scale then enters  the partonic cross section and the 
transport coefficient used  in energy loss calculations~\cite{glv}. 
Larger variations of the initial density of the medium have also been 
studied~\cite{gvlong} but 
they give a less satisfactory description 
of the PHENIX data~\cite{phenix}.

At $p_{\rm T} \geq 2-3$~GeV/c pions are perturbative and exhibit an almost
constant suppression as predicted by the GLV  formalism~\cite{glv,levai}. 
In contrast at moderate high $p_{\rm T}$ the non-perturbative baryon junction
component is unaffected. Thus in central collisions baryon excess at  
intermediate transverse momenta is expected relative to $NN$.
This results in a smaller degree of suppression of inclusive charged 
hadrons which is  $p_{\rm T}$ dependent. At larger 
transverse momenta  baryon production is expected to be dominated 
again by PQCD (as in Fig.1). Therefore baryons are 
expected to show comparable suppression to pions at high 
$p_{\rm T} \gton 5$~GeV/c.
From Fig.~2 we conclude that such physical 
picture is compatible with the data and 
can be easily tested with future data in the $p_{\rm T}>5$~GeV/c range.

\begin{center}
\vspace*{8.7cm}
\includegraphics{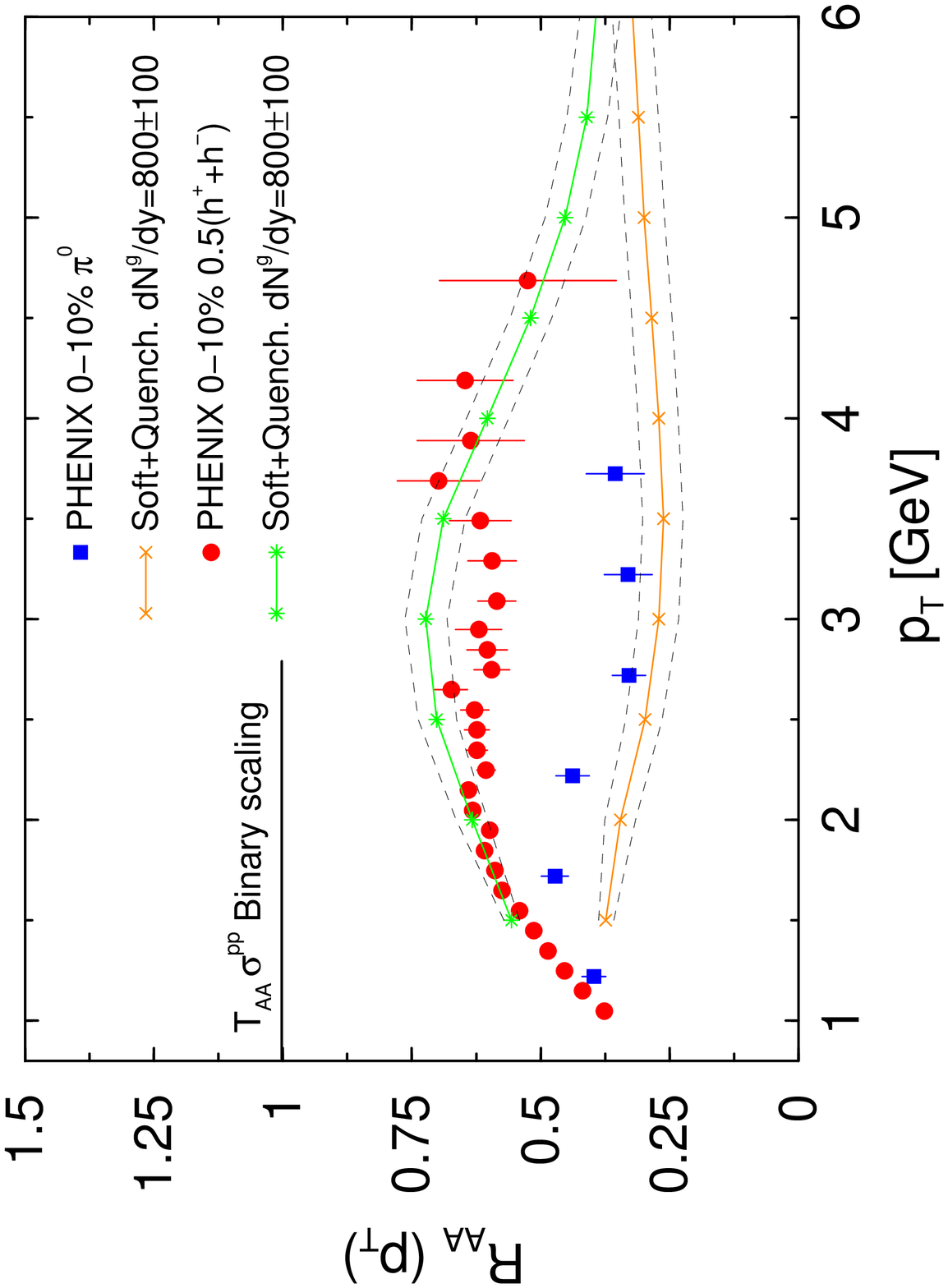}
\vspace{-2.7cm}
\end{center}
\begin{center}
\begin{minipage}[t]{8.5cm}
         { FIG. 2.} {\small  The ratio of charged hadron  and  $\pi^0$ 
multiplicities to the binary collision scaled $\bar{p}p$ result
is shown from~\cite{phenix}.
The curves utilize the GLV quenched hard spectrum and the modified 
soft component Eq.~(\ref{flavslope}). }
\end{minipage}
\end{center}

{\bf The $\bar{p}/\pi^-$ anomaly.}
In our discussion of the $p_{\rm T}$ dependence of  $\bar{p}/\pi^-$
ratio at RHIC we first turn to a simple description based on  boosted 
thermal sources 
\begin{equation}
dN \sim  m_{\rm T}K_1\left( \frac{m_{\rm T} 
\cosh \eta_r}{ T_f} \right) I_0\left( \frac{p_{\rm T} \sinh \eta_r} { T_f}
\right), \;\; 
\label{thermal}
\end{equation}
where $\tanh \eta_r =  v_\perp $. A computation of $R_B(p_{\rm T})$ 
with $T_f=160$~MeV and $v_\perp = 0.6$ is  included in Fig.~3. It 
predicts $\bar{p}/\pi^- \leq 1$ in the intermediate $p_{\rm T}$  window
and grows monotonically with $p_{\rm T}$ to $R_B=2$ which 
is a general feature  of hydrodynamic calculations.  
We note however, that variations of hydrodynamic initial density profiles,
equation of state, and freeze-out criteria as well as 
dissipative effects may also play a role in the
$\bar{p}/\pi^-$ anomaly, since $R_B < 1$ may also occur in hydrodynamics with
late chemical freeze-out~\cite{kolb}
as well as in the  hybrid hydro+UrQMD model due to 
dissipative effects~\cite{dumi}.

As a further test of our model we consider next the 
expected centrality dependence of $R_B$.
The $\bar{p}/\pi^-$ ratio computed for 3 different centralities
in the soft+hard model Eq.~(\ref{hadtot}) is also shown
in Fig.~3. In central collisions the interplay  between the anomalous 
baryon component of $\bar{p}$ and the quenched PQCD component of 
$\pi^-$ leads  to maximum of $R_B$ near  $p_{\rm T}\sim 3-4$~GeV/c. 
At large  $p_{\rm T} \geq  5-6$~GeV/c we predict   
a gradual decrease of $R_B$ {\em below unity}  consistent  with the 
the PQCD baseline calculations (see Fig.~1). In Fig.~3  we have 
included through error bands  the factor of $\sim3$ uncertainty in the 
fragmentation functions into $\bar{p}$ at high $p_{\rm T}$.
The solid and dashed curves reflect the difference between the 
$N_{part}$ and $N_{part}^{4/3}$ scaling of the junction component

In peripheral reactions the size of the interaction region 
as well as the initial density of the medium decrease leading to 
reduction of energy loss. The absence of  quenching
reduces the observability of the anomalous component and 
the $\bar{p}/\pi^-$ ratio may stay  below unity for all $p_{\rm T}$. 
The case of peripheral reactions is hence similar to $\bar{p}p$ collisions.
The experimentally testable  prediction of the model~(\ref{hadtot}) is 
therefore that the maximum of the $R_B=\bar{p}/\pi^- $ ratio decreases
with increasing impact  parameter, decreasing participant number, or 
equivalently decreasing  $dN^{ch}/dy$.

\begin{center}
\vspace*{8.7cm}
\includegraphics{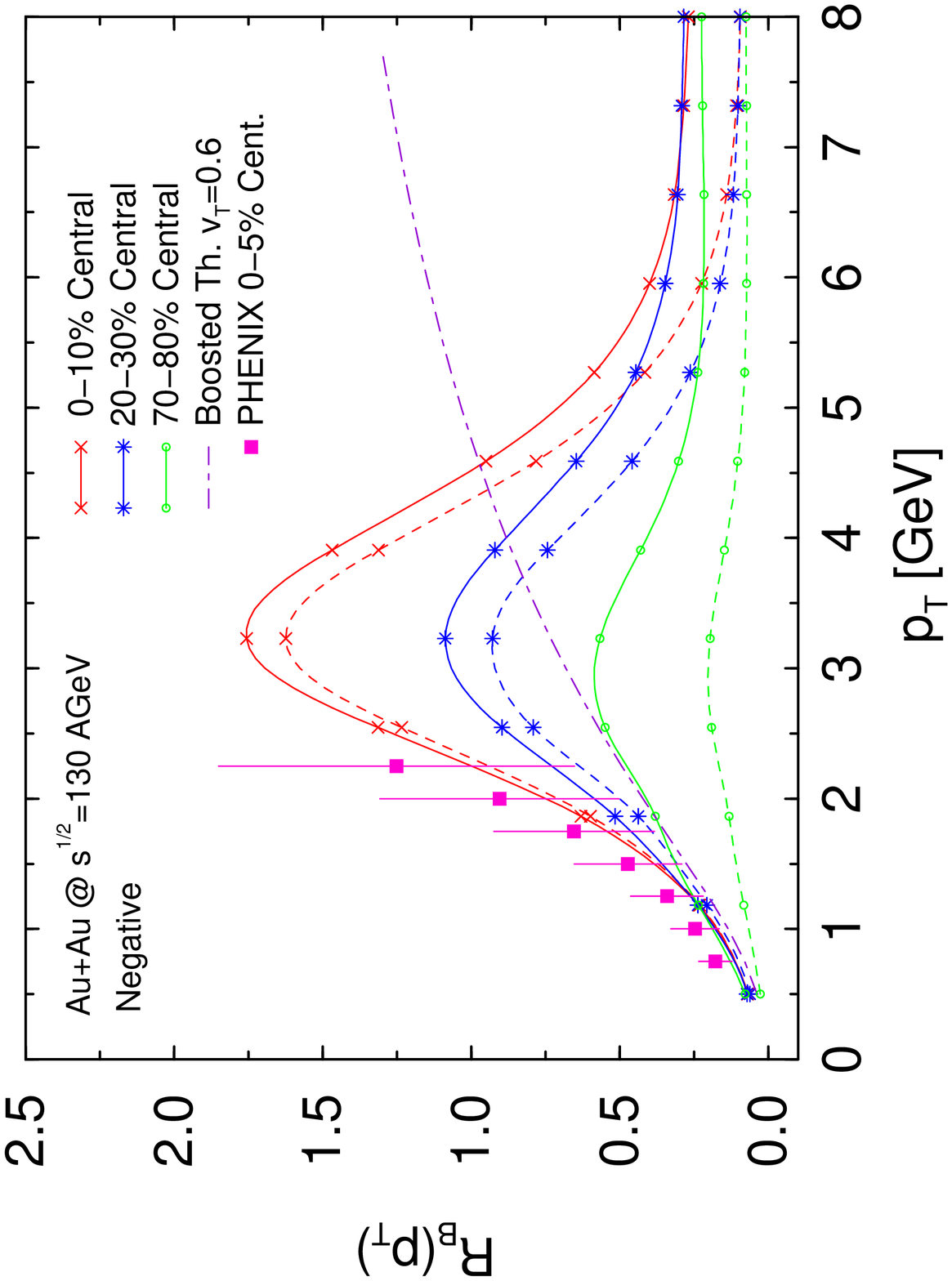}
\vspace{-2.7cm}
\end{center}
\begin{center}
\begin{minipage}[t]{8.5cm}
         { FIG. 3.} {\small  The centrality dependence
of $R_B(p_{\rm T})$ is predicted  for three different centralities. 
Solid (dashed) lines correspond to $A^1$ ($A^{4/3}$) scaling of
the junction component.
The ratio of $\bar{p}$  and $\pi^-$ fits to  PHENIX data 
on central reactions is shown for comparison. A 
boosted thermal source (dashed line) is also shown.}
\end{minipage}
\end{center}

{\bf Conclusions.}
We found that the interplay between jet quenching at RHIC
computed as in~\cite{gvwprl}  using the GVW, GLV 
formalisms~\cite{gvwprl,glv} and 
a postulated novel baryon junction component could account for
the $p_{\rm T}> 2$~GeV/c  $\; \bar{p}/\pi^- \gton 1$ puzzle suggested by
the data. We propose that in central collisions
jet quenching exposes this new baryon junction 
component~\cite{junction,dima96}. This was shown in Fig.~2 to provide a natural
explanation for the smaller effective quenching of 
moderate $p_{\rm T}$ charged particles than $\pi^0$. 
Fig.~3 shows that $R_B$ could reach values well above
unity in a finite moderate $p_{\rm T}$ domain but is predicted to decrease to
well below unity at $p_{\rm T} > 5$ GeV/c in contrast to thermal or 
hydrodynamic models. Whereas the maximum baryon excess, $R_{B \, {\rm max}}$, 
depends on unknown details of the baryon junction component 
the shape of $R_B(p_{\rm T})$
and its specified centrality and transverse momentum
dependence are largely a consequence of the jet quenching phenomenon 
similar to the saturation of $v_2(p_{\rm T})$ reported by STAR~\cite{star}.

Since from $p_{\rm T}$ integrated hadron yields $\bar{p}/p \simeq 0.65$~\cite{phenix,star}, 
the predicted $p_{\rm T}$ dependence of baryon enhancement and subsequent
reduction at high $p_{\rm T}$ may in fact be more
readily observable in the $p/\pi^+$ ratio. 
\begin{center}
\vspace*{8.7cm}
\includegraphics{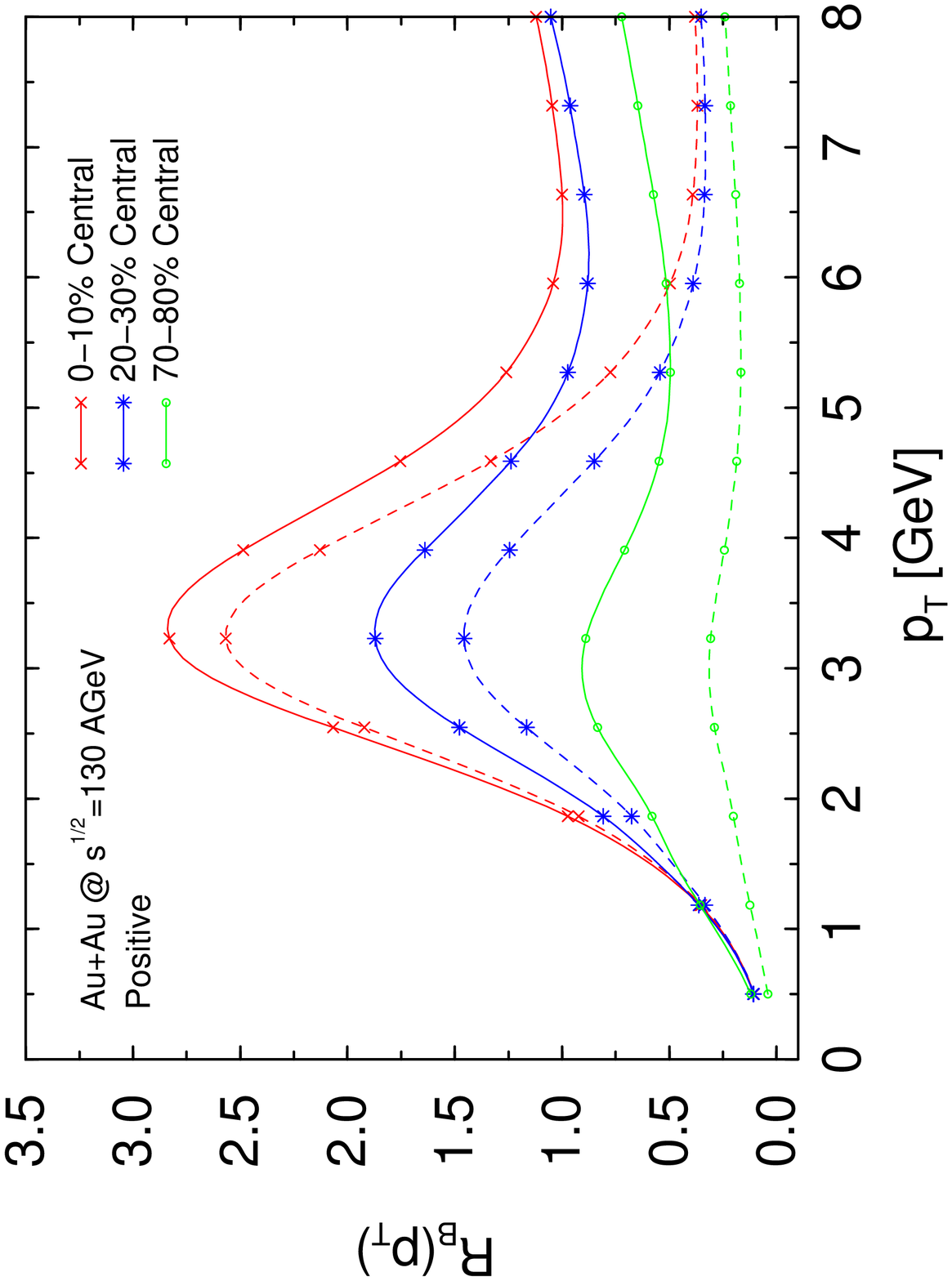}
\vspace{-2.7cm}
\end{center}
\begin{center}
\begin{minipage}[t]{8.5cm}
         { FIG. 4.} {\small  The centrality and $p_{\rm T}$ dependence
of the  $p/\pi^+$ ratio is predicted  for the same centrality classes 
as in Fig.~3. 
The solid (dashed) lines compare the difference between
 $N_{part}$ $(N_{part}^{4/3})$ scaling of the junction component
 and $3(1)\times$ the quenched 
 fragmentation component as in Fig.3.
}
\end{minipage}
\end{center}
We find that the $p/\pi^+$  ratio is typically $\sim 50\%$  
larger than $\bar{p}/\pi^-$ 
as a function of $p_{\rm T}$ at moderate transverse momenta. 
Comparison between Fig.~3 and Fig.~4 shows that for 
$p_{\rm T} \geq 6$~GeV/c the baryon/meson ratio
for positive hadrons predicted here is  $\sim 2-3$  times bigger than 
the corresponding ratio for negative hadrons. This is due to the 
valance quark dominance in the PQCD fragmentation into baryons versus 
antibaryons
at high $p_{\rm T}$ (see also Refs.~\cite{gvlong,wang}). In contrast,
hydrodynamic models with flavor independent freeze-out would predict
an asymptotic ratio $\sim 2$. 

In summary, a novel  test of the baryon junction
hypothesis~\cite{junction,dima96}  for baryon number production and 
transport in ultra-relativistic nuclear collisions has been proposed 
through the predicted 
anomalous dependence of high $p_{\rm T}$ $\bar{p}/\pi^-$ and $p/\pi^+$  
ratios and their centrality dependence. The existing 
PHENIX data~\cite{julia}
strongly motivate the extra effort needed to measure (anti)baryon
spectra into the $p_{\rm T} \gton 5$~GeV/c range as a complement to
the  (anti)baryon number rapidity transport measurements by 
STAR~\cite{nuqm01} and soon BRAHMS~\cite{brahms}. 

{\bf Acknowledgments.} We thank J.~Velkovska, N. Xu and W.~Zajc
for discussions on the RHIC data and A.~Dumitru, U.~Heinz  
and X.-N.~Wang for helpful comments.
This work was supported by the Director, Office of Science, 
Office of High Energy and Nuclear Physics,
Division of Nuclear Physics, of the U.S. Department of Energy
under Contract No. DE-AC03-76SF00098. 
M.G. is also grateful to Collegium Budapest for hospitality and partial 
support during  the completion of this work.

\vfill\eject
\end{multicols}

\vspace{-0.0cm}
\begin{references}
\vspace{-1.cm}

\bibitem{phenix} G.~David, Nucl. Phys. {\bf A698}, 227 (2002); W.~Zajc, 
Nucl. Phys. {\bf A698}, 39 (2002); 
K.~Adcox {\it et al.},  Phys. Rev. Lett. {\bf 88},
 022301 (2002),  [nucl-ex/0109003].   


\bibitem{star} 
K.H.~Ackermann {\it et al.}, Phys. Rev. Lett. {\bf 86}, 402 (2001);
C. Adler {\em et al.}, Phys. Rev. Lett. {\bf 87}, 182301 (2001); 
R.J.~Snellings, Nucl. Phys. {\bf A698}, 193 (2002). 

\bibitem{kolb}P.F.~Kolb, U.~Heinz, P.~Huovinen, K.J.~Eskola, K.~Tuominen,
Nucl. Phys. {\bf A696} 197 (2001).

\bibitem{gvwprl}
M~.Gyulassy, I~.Vitev, X.-N.~Wang, Phys. Rev. Lett. {\bf 86} 2537 (2001). 

\bibitem{pt}  C.~Adler {\em et al.} Phys. Rev. Lett.
{\bf 87} 112303 (2001); A.~Drees, Nucl. Phys. {\bf A698}, 331 (2002).

\bibitem{julia}  J.~Velkovska,  Nucl. Phys. {\bf A698}, 507 (2002); 
K. Adcox {\em et al.}, [PHENIX Collaboration],
nucl-ex/0112006.


\bibitem{nuqm01} N.~Xu, M.~Kaneta, Nucl. Phys. {\bf A698}, 306 (2002);
C.~Adler {\em et al.},  Phys. Rev. Lett. {\bf 87}, 262302 (2001).
 

\bibitem{junction} G.C.~Rossi and G.~Veneziano, Nucl. Phys. {\bf B123} 
507 (1977);  Phys. Rept. {\bf 63} 153 (1980). 

\bibitem{dima96} D.~Kharzeev, Phys. Lett. {\bf B378} 238 (1996);
S.E.~Vance, M.~Gyulassy, X.-N.~Wang,
 Phys. Lett.  {\bf B443} (1998) 45; S.E.~Vance, M.~Gyulassy,
 Phys. Rev. Lett. {\bf 83} (1999) 1735.

\bibitem{gvlong}I.~Vitev, M. Gyulassy and P.~Levai, hep-ph/0109198;
 I.~Vitev, M. Gyulassy, hep-ph/0108045.


\bibitem{quench} 
 X.-N.~Wang, M.~Gyulassy, Phys. Rev. Lett. {\bf 68} 1480 (1992);
Phys.~Rev. {\bf D44} 3501 (1991). 

\bibitem{glv} M.~Gyulassy, P.~L\'evai, I.~Vitev,
Phys. Rev. Lett. {\bf 85} 5535, (2000); 
Nucl. Phys.  {\bf B594} 371  (2001);
Nucl. Phys. {\bf B571} 197 (2000);
Nucl. Phys. {\bf A661} 637c (1999).

\bibitem{levai}
P. Levai, G. Papp, G. Fai and M. Gyulassy, nucl-th/0012017, 
nucl-th/0112062. 


\bibitem{wang}
X.~Wang, Phys. Rev. C {\bf 58} 2321 (1998).

\bibitem{isrua1}
B. Alper {\em et al.}, Phys. Lett. {\bf B44} 521 (1973); 
C. Albajar {\em et al.}, Nucl. Phys. {\bf B335} 261 (1990). 

\bibitem{struct}
M.~Gl\"{u}ck, E.~Reya and W.~Vogt,  Z.~Phys. {\bf C67} 433  (1995);
H.L.~Lai et al., Eur.~Phys.~J. {\bf C12} 375 (2000). 

\bibitem{fragm}
J.~Binnewies, B.A.~Kniehl and G.~Kramer,
Z.~Phys. {\bf C65}, 471 (1995);
B.A.~Kniehl, G.~Kramer and B. Potter,
Nucl.~Phys. {\bf B582}, 514 (2000).  

\bibitem{e735}
T.~Alexopoulos {\it et al.},
Phys. Rev.  {\bf D48} (1993) 984.

\bibitem{gvw2}
M.~Gyulassy, I.~Vitev, X.-N.~Wang, P.~Huovinen, Phys. Lett.~{\bf B526}, 
301 (2002).


\bibitem{cronin}
Y.~Zhang, G.~Fai, G.~Papp, G.~Barnafoldi, P.~Levai, 
Phys. Rev. {\bf C65}, 034903 (2002).

\bibitem{fluc}
M.~Gyulassy, P.~Levai and I.~Vitev, nucl-th/0112071.

\bibitem{Topor02}
V.~Topor Pop {\em et al.}, in preparation. 

\bibitem{rischke}
D.H.~Rischke, M.~Gyulassy, Nucl.~Phys.~{\bf A608}, 479 (1996).


\bibitem{dumi}
S.A.~Bass, A.~Dumitru, Phys.~Rev.~{\bf C61}, 064909 (2000). 


\bibitem{phobos} QM2001, 
B.B.~Back {\it et al.}, Phys. Rev. Lett. {\bf 85}, 3100 (2000);
K. Adcox {\em et al.}, Phys. Rev. Lett. {\bf 86}, 3500 (2001). 

\bibitem{exdata}
D.~Antreasyan {\em et al.}, Phys. Rev. {\bf D19}, 764 (1979).

\bibitem{cern} I.~Vitev, Talk at ``2001 CERN Workshop on
 Hard Probes in Heavy Ion Collisions at the LHC'',
wwwth.cern.ch/lhcworkshop/TALK\_oct01/vitev/vitev.pdf

\bibitem{ekrt}
K.J.~Eskola, K.~Kajantie, P.V.~Ruuskanen, K.~Tuominen,
Nucl. Phys. {\bf B570}, 379 (2000). 

\bibitem{brahms} I.G.~Bearden {\em et al.}, [BRAHMS Collaboration], 
nucl-ex/0112001. 

\end{references}
\end{document}